\documentclass[dvips, usenatbib, fleqn]{mn2e}
\usepackage{mathptmx}
\usepackage[dvips]{graphicx}
\usepackage{amsmath}
\usepackage{mn2e_fixbib}
\usepackage{widetable}
\usepackage{booktabs}
\usepackage{caption}
\usepackage{balance}
\begin{document}

\title[Core radii and common-envelope evolution]{Core radii and common-envelope evolution}

\author[P.~D.~Hall and C.~A.~Tout]{%
  Philip~D.~Hall\thanks{E-mail:~pdh37@ast.cam.ac.uk} and Christopher~A.~Tout\\
  Institute of Astronomy, University of Cambridge, The Observatories, Madingley Road, Cambridge CB3 0HA, UK \\
  }
\date{Accepted 2014 August 14. Received 2014 August 4; in original form 2014 June 16}
\volume{444}
\pagerange{3209--3219}
\pubyear{2014}

\maketitle

\label{firstpage}

\begin{abstract}
Many classes of objects and events are thought to form in binary star systems after a phase in which a core and companion spiral to smaller separation inside a common envelope (CE).
Such a phase can end with the merging of the two stars or with the ejection of the envelope to leave a surviving binary system.
The outcome is usually predicted by calculating the separation to which the stars must spiral to eject the envelope, assuming that the ratio of the core--envelope binding energy to the change in orbital energy is equal to a constant efficiency factor $\alpha$.
If either object would overfill its Roche lobe at this end-of-CE separation, then the stars are assumed to merge.
It is unclear what critical radius should be compared to the end-of-CE Roche lobe for stars which have developed cores before the start of a CE phase.
After improving the core radius formulae in the widely used \textsc{bse} rapid evolution code, we compare the properties of populations in which the critical radius is chosen to be the pre-CE core radius or the post-CE stripped remnant radius.
Our improvements to the core radius formulae and the uncertainty in the critical radius significantly affect the rates of merging in CE phases of most types.
We find the types of systems for which these changes are most important.
\end{abstract}


\section{Introduction}\label{sec:intro}
Classes of short-period binary star systems containing compact objects such as cataclysmic variables (CVs), low-mass X-ray binary systems, white dwarf--white dwarf binary systems and the binary central stars of planetary nebulae are important to many astrophysical phenomena.
The past evolution of all these objects included a phase of common-envelope (CE) evolution in which one component was engulfed by the other.
A CE phase can begin when a giant star in a relatively long-period binary system begins Roche lobe overflow.
In such a situation, the mass-loss rate from the giant can rapidly increase to a rate at which the companion becomes immersed in the common giant envelope and spirals towards the giant's core, depositing energy and angular momentum in the envelope.
This continues until the components merge to form a single star or until the envelope is ejected, to leave the remnant of the enveloper and its companion in a shorter period binary system \citep{Pac76:73}. 
In this way, systems of the types mentioned are formed as the core of the enveloper is exposed and the orbital separation is decreased.

Our understanding of CE evolution remains uncertain \citep{Iva13:59}.
There are currently no realistic models: the most recent three-dimensional hydrodynamical models of CE evolution end after an initially rapid phase of spiral-in and before most of the envelope is unbound \citep{Pas12:52, Ric12:74}.
Thus, we are unable to relate the pre- and post-CE configurations using such models.
Predictions of the outcome of CE evolution in both detailed and rapid evolution codes therefore usually use a simple recipe such as the $\alpha$ prescription.
There are several variations of this prescription but a representative version is that described by \citet*{Hur02:897} and used in their \textsc{bse} rapid evolution code.
If a binary system satisfies the conditions for the onset of a CE phase, and only one component, the enveloper, has developed a core, then the binding energy $E_{\rmn{bind}}$ is computed from
\begin{equation}\label{eq:ebind}
  E_{\rmn{bind}} 
= -\frac{G M M_{\rmn{env}}}{\lambda R},
\end{equation}
where $G$ is the gravitational constant, $M$ is the total mass of the enveloper, $M_{\rmn{env}}$ is the envelope mass, $R$ is the radius and $\lambda$ is a constant (usually taken to be about $0.5$) or computed from detailed models \citetext{\citealp{Xu10:114}; \citealp*{Lov11:49}}.
It is implicitly assumed that any binding energy interior to the CE remnant mass $M_{\rmn{rem}} = M - M_{\rmn{env}}$ is unchanged when the envelope of mass $M_{\rmn{env}}$ is removed.
It is assumed that the binding energy and the difference in orbital energy between pre- and post-CE configurations $\Delta E_{\rmn{orb}}$ are related by an efficiency factor $\alpha = E_{\rmn{bind}} / \Delta E_{\rmn{orb}}$, which allows computation of the separation to which the cores must spiral if sufficient energy is to be released to eject the envelope, assuming that only $M_{\rmn{rem}}$ from the original star remains and that the companion is unchanged.

Our focus is the condition for survival as a binary system or, equivalently, the condition to avoid merging.
The usual condition in rapid evolution codes, such as \textsc{bse}, is that if either the companion or the CE remnant would overfill its Roche lobe at the calculated separation, then the cores merge.
Otherwise the envelope is ejected and the system survives as binary at that separation.
The radius of the companion is known on the assumption that it is mostly unaffected by CE evolution.
Thus, it is reasonable to check whether the companion overfills its Roche lobe in an orbit with the putative final separation.
However, in some cases, the core CE remnant is likely to be larger than its companion and closer to filling its Roche lobe.
In these cases, it is unclear what radius should be compared to the Roche lobe radius.

Properly, the models produced by a rapid evolution code such as \textsc{bse} should be viewed as a simplified version of a more complex situation.
Hence, we should look to more detailed prescriptions to find the appropriate comparison radius.
However, there are several possible detailed treatments and it is difficult to see how they should be simplified for this purpose.
For example, in the CE prescription of \citet{Ibe86:742, Ibe89:505}, the radius of the remnant is in thermal equilibrium and its envelope mass is chosen such that it just fits inside its end-of-CE Roche lobe.
The implied condition for merging in this case might be implemented in a rapid evolution code by requiring that the completely naked thermal-equilibrium remnant of mass almost equal to the core mass just fits inside its Roche lobe.
However, other prescriptions have been suggested based on computing adiabatic mass-loss sequences \citep{Del10:L28,Ge10:724,Iva11:76}.
These prescriptions cannot be neatly implemented in \textsc{bse} (or any rapid evolution code) because they depend on the detailed behaviour of stars under extremely rapid, or adiabatic mass-loss, which is not included in \textsc{bse}.
In these CE prescriptions, the components merge if there is unstable mass transfer in the post-ejection phase in which the enveloper is left out of thermal equilibrium.
Currently in \textsc{bse} the material and radius are fixed to be those of the completely stripped core.
The enveloper is assumed to be stripped of its entire envelope in a CE phase and an estimate of the pre-CE core radius is compared to the Roche lobe radius.
To take the critical radius to be the pre-CE core radius could be interpreted as implying a choice of a completely dynamical CE phase which is rapid enough that the structure of the enveloper does not significantly change.
Because of this uncertainty about what is the best choice for the critical radius, we investigate for which types of CE phase the critical radius is particularly important.
We model two choices for the critical radius in \textsc{bse}.
These are the extremes of the core radius at the beginning of the CE phase, as currently implemented, or the radius of the naked CE remnant, stripped of the entire envelope.

If the critical radius is the pre-CE core radius, then we need reliable estimates of this quantity.
Because of this, we first derive an improved prescription for the core radii from a set of detailed stellar models.
We then investigate how important the CE critical radius is to the prediction of properties of populations of binary star systems.
To this end, in Section~\ref{sec:codes} we describe a set of single-star models, computed with the Cambridge stellar-evolution code \textsc{stars}.
In Section~\ref{sec:formulae}, we compare the core radii of these detailed models for different definitions of the core, with \textsc{bse} models and derive improved formulae for the core radii.
In Section~\ref{sec:popsyn}, we describe how we compute the properties of model populations of binary star systems.
In Section~\ref{sec:results}, we compare the effect of choosing the new formulae for the core radius with the alternative of the zero-age remnant in the synthetic populations.
We also compare the output of the \textsc{bse} and new formulae for the core radii.
We identify which types of CE phase particularly depend on the critical radius and conclude in Section~\ref{sec:conclusion}.

\section{Evolution models}\label{sec:codes}
For the rapid computation of the evolution of binary systems required for population synthesis, we use the \textsc{bse} code.
To find the core radii of evolved stars, we make a set of hydrogen-rich and helium-rich models for different masses and metallicities with the Cambridge stellar-evolution code.

\subsection{The \textsc{bse} code}
The \textsc{bse} code was first described by \citet*{Hur00:543} and \citet{Hur02:897}.
It uses analytic formulae and simple prescriptions to compute a model of the evolution of a binary system in a much shorter time than a detailed evolution code.
In its simplest application, the code models the evolution of binary star systems of given zero-age parameters and yields the eccentricity, separation, and masses of the two components, their luminosities, effective temperatures, core masses and radii for any system age.
Since the original work, the code has been updated as described by \citet{Kie06:369} and \citet{Kie08:388}.
Most significant for our interest in CE evolution is the implementation of a variable $\lambda$ (equation~\ref{eq:ebind}) based on fits to detailed models.
We also use the new prescription for the post-CE spins of the components of a surviving binary system.
Previously, the spins of the stars were set to corotate with the orbit at the end of the CE phase.
Now we set the spins to their pre-CE values, which can then synchronize by tidal interaction.
There are many other parameters, such as the $\alpha$ efficiency parameter, which affect the evolution.
For direct comparison we set these as in table 3 of the work by \citet{Hur02:897}.

The \textsc{bse} code includes an algorithm for computation of the properties of a \emph{single} star of given mass, metallicity and age.
Properties calculated are the surface luminosity, radius, core mass and core radius.
This algorithm, including the core radius prescriptions, was devised by \citet{Hur00:543} from models computed by \citet{Pol98:525}.
Other authors of rapid binary star evolution algorithms have used the formulae in their work.
Examples of such codes include \textsc{binary\_c} \citep{Izz09:1359}, \textsc{biseps} \citep{Wil02:1004}, \textsc{seba} \citetext{\citealp{Nel01:939}; \citealp*{Too12:70}} and \textsc{startrack} \citep{Bel08:223}.
The formulae and uncertainties discussed here are relevant to users of these codes too.

\subsection{The stellar-evolution code \textsc{stars}}
We compute detailed single-star models with the version of the Cambridge stellar-evolution code \textsc{stars} described by \citet{Sta09:1699} which has descended from that written by \citet{Egg71:351, Egg72:361} and updated by \citet{Pol95:964}. 
It is available at http://www.ast.cam.ac.uk/$\sim$stars.
Model sequences produced by the code satisfy a standard set of one-dimensional quasi-static stellar-evolution equations with meshpoints distributed in a non-Lagrangian mesh \citep{Egg71:351}. 
The convective mixing-length \citep{Boh58:108} parameter $\alpha_{\rmn{MLT}}=2$ and convective overshooting is included with a parameter $\delta_{\rmn{ov}}=0.12$ \citep*{Sch97:696} for consistency with the work of \citet*{Hur00:543}. 
These convective parameters are consistent with observations of the Sun, open clusters and spectroscopic binary systems \citep{Pol98:525} when we assume a metallicity $Z=0.02$.

The equation of state and other thermodynamic quantities are described by \citet*{Egg73:325} and \citet{Pol95:964}. 
The radiative opacity is that of the OPAL collaboration \citep{Igl96:943}, supplemented by the molecular opacities of \citet{Ale94:879} and \citet{Fer05:585} for low temperatures and of \citet{Buc76:440} for pure electron scattering at high temperatures. 
The electron conduction opacity is taken from the work of \citet{Hub69:18} and \citet{Can70:641}. 
The construction of the opacity tables and their inclusion in the code was described by \citet{Eld04:201}. 
The nuclear-reaction rates are those of \citet{Cau88:283} and the NACRE collaboration \citep{Ang99:3}, as described by \citet{Eld04:87} and \citet{Sta05:375}. 
The enhancement of reaction rates by electron screening is included according to \citet{Grab73:457}.
The rates of energy loss in neutrinos are due to \citet{Ito89:354, Ito92:622} for the photo/pair and plasma processes, respectively, and \citet{Ito83:858} and \citet*{Mun87:708} for the bremsstrahlung process.

\subsection{Normal stars}
We compute the evolution of normal hydrogen-rich stars from the zero-age main sequence (ZAMS) to envelope exhaustion, degenerate carbon ignition or the end of carbon burning.
We compute models with seven zero-age metallicities, $Z=0.0001$, $0.0003$, $0.001$, $0.004$, $0.01$, $0.02$ and $0.03$ to cover the same range as \textsc{bse}.
At the ZAMS the models are in complete thermal equilibrium with uniform abundance profiles.
We choose zero-age helium mass fractions $Y=0.24+2Z$, so that there is a constant rate of helium to metal enhancement, $\Delta Y/\Delta Z=2$  from the primordial abundance $(Y_{\rmn{p}}, Z_{\rmn{p}})=(0.24, 0)$, calibrated to near-solar abundance $(0.28, 0.02)$. 
For $Z=0.001$, this gives $Y=0.242$.
The relative abundances of the various metals in the initial configuration are constant throughout all models, scaled to the solar abundances given by \citet{And89:197}.
In our models only the mass fractions of ${}^1\rmn{H}$, ${}^3\rmn{He}$, ${}^4\rmn{He}$, ${}^{12}\rmn{C}$, ${}^{14}\rmn{N}$, ${}^{16}\rmn{O}$ and ${}^{20}\rmn{Ne}$ are evolved by convective mixing and nuclear reactions.
These nuclides are sufficient because they determine the structure of the stars of interest here.
For each $Z$, we compute 25 models with masses between $0.5$ and $50\,\rmn{M}_{\sun}$, evenly spaced in $\log M$.
There are thus $175$ stars in our set of detailed models which cover the range of parameters allowed by \textsc{bse}.
The stars evolve at constant mass.

Because we are unable to converge models during the helium-core flash in low-mass stars, we follow the recipe used by others to produce zero-age horizontal-branch models quiescently burning helium with the same composition profile as at the onset of the flash \citep{Pol98:525}.
During the asymptotic giant branch (AGB) phase we do not resolve thermal pulses \citep{Egg73:279}.

\subsection{Naked helium stars}
We also compute the evolution of naked helium stars from the zero-age helium main sequence (MS), which we take to be an initially homogeneous and thermal equilibrium configuration.
We only compute models with $Z=0.02$ to be consistent with \textsc{bse}.
We set the abundances of metals to be the same as normal ZAMS stars and the abundance of helium-3 to zero.
We compute $25$ models with masses between $0.32$ and $10\,\rmn{M}_{\sun}$, evenly spaced in $\log M$.

\subsection{Definition of the core boundary}
We consider three definitions for the hydrogen-exhausted core (for normal hydrogen-rich stars) and the helium-exhausted core (for naked helium stars).
Our first definition for the core boundary is the point at which the hydrogen or helium abundance by mass fraction $X$ or $Y=0.1$.
This definition is chosen for consistency with the core-mass formulae used in \textsc{bse}.
This normally gives a point which is corresponding to the idea of the outer boundary of the core at the base of the burning shell in phases when such a shell is active.
It gives an outer boundary to the region where the hydrogen or helium abundance is low.
We also compute the enclosed mass and radius where $X$ or $Y$ is $0.01$ and $0.001$.
These are reasonable alternative definitions of the core boundary.
For each type of star (phase, zero-age mass and composition), we compare our core radii with these alternative definitions to assess how well defined the core is in terms of mass and radius.
Generally, the mass of an exhausted core is a much better defined notion than its radius.
For example, for low-mass red giant branch (RGB) stars the enclosed masses at $X=0.1$, $0.01$ and $0.001$ are the same within $0.1$ per cent but the radii at the same points differ by more than $10$ per cent.
Although the core radius is not always well defined, we show in the next section that this uncertainty is less than the discrepancy between the current \textsc{bse} formulae and what we find in detailed models.

Note that we do not claim that $X=0.1$ is the point to which matter is removed during a CE phase, the remnant--envelope boundary or bifurcation point, as it is sometimes called \citep{Tau01:170}.
The remnant--envelope boundary is normally considered to be further away from the centre of the star.
Our choice of $X=0.1$ is a consistent definition for the exhausted core boundary and can be used as a basis to consider possible conditions for merging during a CE phase.

\section{Core radius fitting formulae}\label{sec:formulae}
We compare the core radii of \textsc{bse} models with those in detailed models and improve the former where necessary.
In the \textsc{bse} scheme, the core radius is set to a multiple of the radius of the remnant if matter were to be removed to the core boundary.
In \textsc{bse}, the remnants are white dwarfs in the case of stars with degenerate cores and naked helium stars in the case of stars with non-degenerate cores.
We follow the same approach here.
For reference, the \textsc{bse} formula for the radius of a cold white dwarf with mass $M$ is
\begin{equation}\label{eq:RWD}
  \frac{R_{\rmn{WD}}(M)}{\rmn{R}_{\sun}} 
= 0.0115 \sqrt{\left( \frac{M_{\rmn{Ch}}}{M} \right)^{2/3} - 
\left( \frac{M}{M_{\rmn{Ch}}} \right)^{2/3} },
\end{equation}
where the Chandrasekhar mass $M_{\rmn{Ch}} = 1.44\,\rmn{M}_{\sun}$, while the radius of a ZAMS helium star with mass $M$ is
\begin{equation}
  \frac{R_{\rmn{He}}(M)}{\rmn{R}_{\sun}} 
= \frac{0.2391 (M/\rmn{M}_{\sun})^{4.6}}{(M/\rmn{M}_{\sun})^4 + 0.162 (M/\rmn{M}_{\sun})^3 + 0.0065}
\end{equation}
\citep{Hur00:543}.

Our detailed models are analysed in terms of the evolutionary phases defined by \citet{Hur00:543, Hur02:897} and the types summarized in Table~\ref{tab:types}.
We also use their definition of low-, intermediate- and high-mass stars.
For a given metallicity, $M_{\rmn{HeF}}(Z)$ is the maximum mass of a constant mass star which ignites helium in its core under degenerate conditions.
Low-mass stars have masses less than this.
Intermediate-mass stars are more massive than $M_{\rmn{HeF}}(Z)$ and ignite helium when on the giant branch.
High-mass stars are more massive than $M_{\rmn{HeF}}(Z)$ and ignite helium before reaching the giant branch.

\begin{table}
\centering
\caption{Stellar types defined by \citet{Hur02:897}.}
\label{tab:types}
\small
\begin{tabular}{rl} 
\toprule
$k$ & Description \\
\midrule
0 & Main sequence (MS, $M < 0.7\,\rmn{M}_{\sun}$) \\
1 & Main sequence (MS, $M \geq 0.7\,\rmn{M}_{\sun}$) \\
2 & Hertzsprung gap (HG) \\
3 & Red giant branch (RGB) \\
4 & Core helium burning (CHeB) \\
5 & Early asymptotic giant branch (EAGB) \\
6 & Thermally pulsing AGB (TPAGB) \\
7 & Naked helium star MS (HeMS) \\
8 & Naked helium star Hertzsprung gap (HeHG) \\
9 & Naked helium star giant branch (HeGB) \\
10 & Helium white dwarf (HeWD) \\
11 & Carbon--oxygen white dwarf (COWD) \\
12 & Oxygen--neon white dwarf (ONeWD) \\
13 & Neutron star (NS) \\
14 & Black hole (BH) \\
15 & Massless remnant \\
\bottomrule
\end{tabular}
\end{table}

\subsection{Hertzsprung gap and RGB stars}
Once core hydrogen burning ceases at the end of MS evolution, a helium core develops.
MS stars less massive than about $1.2\,\rmn{M}_{\sun}$ burn hydrogen radiatively so the core grows from $m=0$ during the MS.
More massive MS stars burn hydrogen convectively and so the core forms with finite mass.

\subsubsection{Low-mass stars}

\begin{figure}
  \includegraphics[width=84mm]{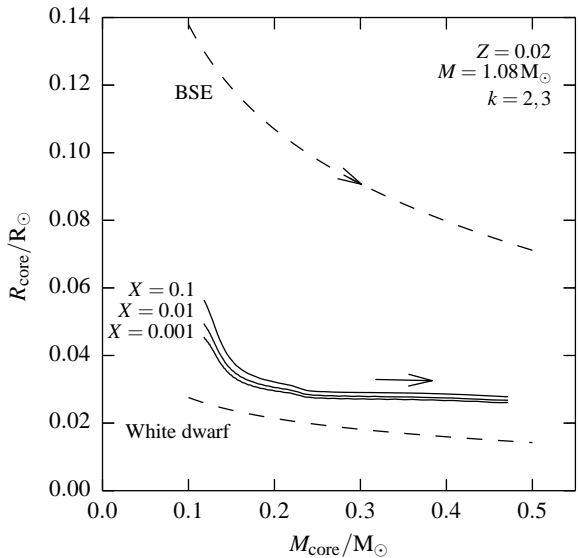}
  \caption{
The evolution of a $1.08\,\rmn{M}_{\sun}$, $Z=0.02$ star in the core mass--core radius plane, for different definitions of the core boundary.
We compare the radii with the \textsc{bse} formulae for the core radius and the radius of white dwarfs.
The \textsc{bse} core radii are too large by at least $130$ per cent.}
  \label{fig:RGB-lowmass-eg}
\end{figure}

\begin{figure}
  \includegraphics[width=84mm]{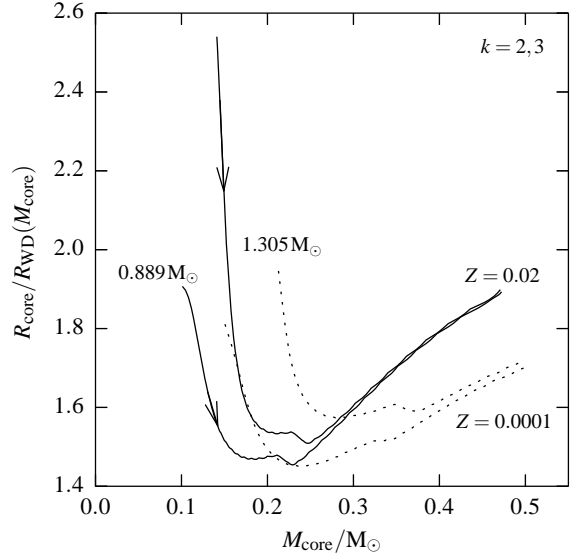}
  \caption{
The ratio of core radius to the radius of a cold white dwarf of mass equal to the core mass, for low-mass RGB stars with masses $0.889$ and $1.305\,\rmn{M}_{\sun}$ and metallicities $0.0001$ and $0.02$.
\textsc{bse} sets this ratio to be $5$.}
  \label{fig:RGB-lowmass-ratio}
\end{figure}

In \textsc{bse}, a Hertzsprung gap (HG) or RGB star with $M<M_{\rmn{HeF}}$ has core radius which is five times the radius of a cold white dwarf with mass equal to the core mass.
This choice was originally made as the simplest fit to a small number of models.
The formula, like others in \textsc{bse} was always open to improvement.
\mbox{Fig.~\ref{fig:RGB-lowmass-eg}} shows the evolution of our $Z=0.02$, $1.08\,\rmn{M}_{\sun}$ model in these phases for different definitions of the core radius.
During these phases the core contracts while its mass grows by hydrogen-shell burning.
For core masses exceeding about $0.25\,\rmn{M}_{\sun}$, during the RGB phase, the core radius definitions agree to better than $10$ per cent.
The core radius is less well defined in the HG phase.
Also shown in \mbox{Fig.~\ref{fig:RGB-lowmass-eg}} is the \textsc{bse} core radius which needs improving because it is always at least $130$ per cent larger than in the detailed model.

\mbox{Fig.~\ref{fig:RGB-lowmass-ratio}} shows the ratio of core radius (radius at $X=0.01$) to white dwarf radius of four low-mass stars, with masses $M=0.889$ and $1.31\,\rmn{M}_{\sun}$ and metallicities $Z=0.0001$ and $0.02$.
This figure shows that, contrary to \textsc{bse}, the relation between core radius and core mass depends on both mass and metallicity; this is particularly clear during the HG phase, before the core is well defined and degenerate.
Therefore, an accurate fit would require fitting to both mass and metallicity.
However, a complicated mass-dependent fit for the HG phase is not justified here because the HG phase is short-lived.
Indeed, our models are not sufficiently close to the \textsc{bse} fits.
For example, the core masses at the critical stages of the start of the HG and RGB differ.
In this investigation, we wish to improve the core radius fit for the RGB where a good fit in terms of core mass and $Z$ exists.
For low-mass stars, the RGB is more important than the HG for CE evolution because during the RGB phase stars grow significantly more and the core is better defined.
Binary star systems in which HG stars overfill their Roche lobes really ought to be modelled with a detailed code rather than \textsc{bse}.
\mbox{Fig.~\ref{fig:RGB-lowmass-ratio}} shows that for RGB stars, the core radius to white-dwarf radius ratio is approximately independent of mass $M$ and depends on metallicity $Z$ alone.
To find a formula for the core radius at a given core mass and $Z$, we fit a third-order polynomial in $M_{\rmn{core}}/\mathrm{M}_{\sun}$ to the ratio of core radius to white-dwarf radius.
We fit this polynomial to a representative low-mass model, the mid-point in mass, for the HG and RGB phases at each $Z$.
We then fit a polynomial in $\log_{10}(Z/0.02)$ to each of the polynomial coefficients so that we can interpolate for intermediate metallicity values.
The coefficients in these polynomials are given in \mbox{Table~\ref{tab:LM-HG-RGB-ratio}} of Appendix A.
The resulting fits give much more accurate estimates than \textsc{bse} for the core radius in these phases.
With the new formula for core radius, the core radii are accurate to better than $20$ per cent.
This is a big improvement on the sometimes $300$ per cent error with the standard \textsc{bse} formulae.

\subsubsection{Intermediate-mass stars}\label{sec:intmass}

\begin{figure}
  \includegraphics[width=84mm]{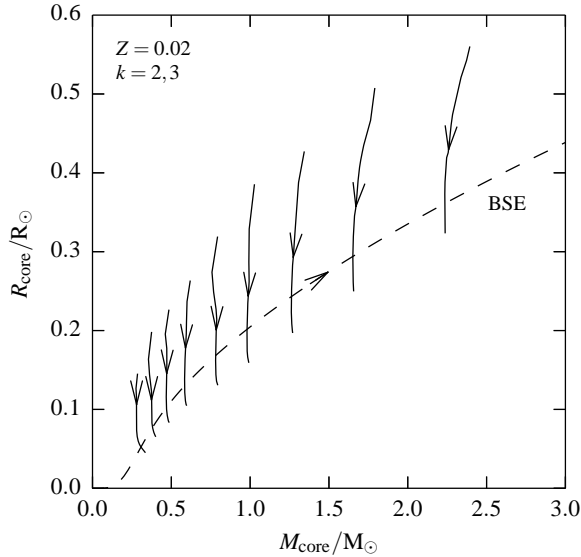}
  \caption{
Evolution of intermediate-mass HG and RGB stars in the core mass--core radius plane from the end of core hydrogen burning to the ignition of CHeB.
The curves from left to right are for masses $2.32$, $2.81$, $3.41$, $4.13$, $5.00$, $6.06$, $7.34$, $8.89$ and $10.77\,\rmn{M}_{\sun}$.
We compare the radii with the \textsc{bse} formulae for the core radius.}
  \label{fig:RGB-intmass-all}
\end{figure}

\begin{figure}
  \includegraphics[width=84mm]{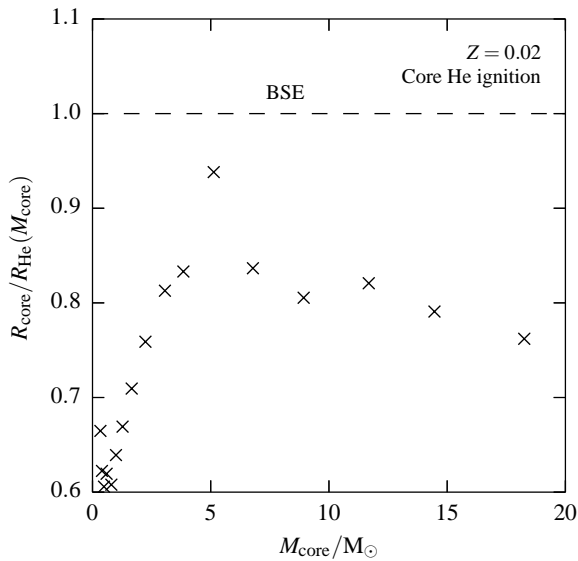}
  \caption{
Ratio of core radius at point of core helium ignition to the radius of a zero-age helium star of mass equal to the core mass for $Z=0.02$ stars.
\textsc{bse} sets this ratio to be $1$.}
  \label{fig:RGB-intmass-ratio}
\end{figure}

In \textsc{bse}, an intermediate-mass HG or RGB star has core radius equal to the radius of a ZAMS helium star of mass equal to the core mass, $R_{\rmn{core}} = R_{\rmn{He}}(M_{\rmn{core}})$.
As is the case for low-mass stars, the core mass and radius are less well defined in the HG than in the RGB phase.
Fig.~\ref{fig:RGB-intmass-all} shows the evolution in the core mass--core radius plane for $Z=0.02$ intermediate-mass stars in the HG and RGB phases, from the end of core hydrogen burning to the ignition of core He burning (CHeB).
In each case, the core contracts at nearly constant core mass, and then the core mass increases by a small amount while the core radius decreases by a small amount before helium is ignited at the tip of the giant branch.
There is a slight unphysical decrease in the core mass shown in the figure as numerical diffusion blurs the composition boundary.
We check the results in models with many more meshpoints and find that the models at later times are not significantly affected.
This behaviour is consistent with a rapid excursion, from the end of the MS across the HG to the RGB, during which the core acquires some electron degeneracy support on a time-scale much less than the nuclear time-scale on which the core can grow.

In \textsc{bse} models of intermediate-mass stars, the core radius does not decrease when the core mass increases by the small amount in the HG and RGB phases -- the core radius increases along the dashed line in Fig.~\ref{fig:RGB-intmass-all}.
It is difficult to correct this behaviour because the contraction of the core during the HG phase depends on the total mass $M$ and would therefore require a complicated fit.
Again, we note that \textsc{bse} should not be used to make quantitative conclusions on stars which transfer mass during their HG phase.
We could attempt to correct for the discrepancy in the core radius at the point of core helium ignition (the tip of the RGB), near which Roche lobe overflow is more likely to begin because the RGB star is largest.
Fig.~\ref{fig:RGB-intmass-ratio} shows the ratio of the core radius at this stage to the radius of a naked helium star $R_{\rmn{He}}(M_{\rmn{core}})$ for intermediate- and high-mass stars.
The core radius is usually less than, but not much less than, the \textsc{bse} CE remnant radius during these phases.
Hence, we do not improve the current prescription for the core radius because it would not be useful for \textsc{bse} which immediately replaces the RGB core with a naked helium star when a CE terminates.
If the CE critical radius were taken to be the pre-CE core radius in the prescription for merging and the core radius is smaller than the naked remnant of the same mass, \textsc{bse} could make a post-CE system in which the CE remnant immediately overfills its Roche lobe.
Therefore, for this work, there is little advantage of improving the core radii in this case.
The results would be the same as taking the critical radius to be the naked remnant as the condition for merging, because the naked helium star would fill its Roche lobe.

\begin{figure}
  \includegraphics[width=84mm]{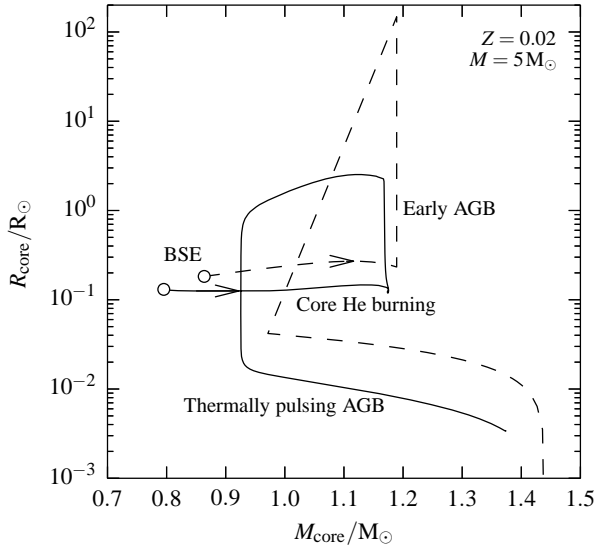}
  \caption{
Evolution of a $Z=0.02$, $5\,\rmn{M}_{\sun}$ star (solid line) and corresponding \textsc{bse} model (dashed line) in the core mass--core radius plane.
The \textsc{bse} core is generally smaller than in the detailed model for the phases in which the star has a non-degenerate helium core.
When it reaches the TPAGB, the core becomes too large.}
  \label{fig:RGB-intmass-eg}
\end{figure}

That the pre-CE core is smaller than the CE remnant is also true for the later CHeB and early asymptotic giant branch (EAGB) phases for stars of all masses.
In these phases, \textsc{bse} also sets the core radius to be that of a naked helium star of appropriate mass and evolutionary state.
The \textsc{bse} core radii could be improved for all these types by multiplication with a core-mass dependent factor for each $Z$.
This is demonstrated in Fig.~\ref{fig:RGB-intmass-eg}, which shows the evolution of a $Z=0.02$, $5\,\rmn{M}_{\sun}$ star in the core mass--core radius plane.
The general shape of the path followed by the \textsc{bse} model is consistent with the detailed model, the core radius increases during the CHeB phases,  but the core is always smaller than a naked helium star of the same mass and evolutionary state until late in the EAGB phase and the thermally pulsing asymptotic giant branch (TPAGB) phase.
Second dredge-up is instantaneous in \textsc{bse} so that there is a difference in the shape of the path in this phase -- a sharp straight line decrease in core mass and radius for the \textsc{bse} model and a smoother decrease for the detailed model.
This cannot be improved without modifying the treatment of second dredge-up in \textsc{bse}.
The core radii could be improved and a smooth core radius change maintained by fitting a single factor, to the start of the CHeB phase which is then multiplied by the current \textsc{bse} core radius during the phases in which the star has a non-degenerate helium core to give a better estimate.
Again, we note that for \textsc{bse} purposes, there is no great advantage of improving the core radii because the outcome of CE evolution would be unaffected in our scheme.
As it is, for intermediate-mass stars in the HG and RGB phases, \textsc{bse} gives a core radius which is too large by about $20$ per cent (Fig.~\ref{fig:RGB-intmass-ratio}).
The same argument applies to high-mass HG stars, CHeB stars and EAGB stars.
In all of these phases, the core radius is taken to be that of an appropriately evolved naked helium star.

\subsection{TPAGB stars}

\begin{figure}
  \includegraphics[width=84mm]{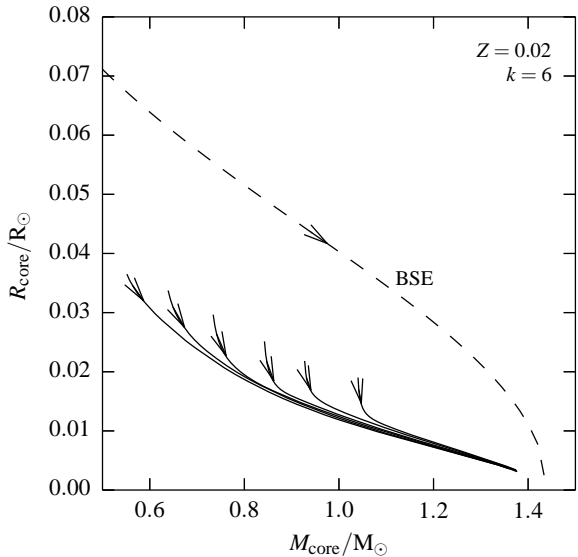}
  \caption{
The evolution in the core mass--core radius plane of $Z=0.02$ intermediate-mass models in the TPAGB phase.
The curves from left to right are for masses $2.32$, $2.81$, $3.41$, $4.13$, $5.00$ and $6.06\,\rmn{M}_{\sun}$.
The \textsc{bse} core mass--core radius relation is marked.
}
  \label{fig:TPAGB-all}
\end{figure}

\begin{figure}
  \includegraphics[width=84mm]{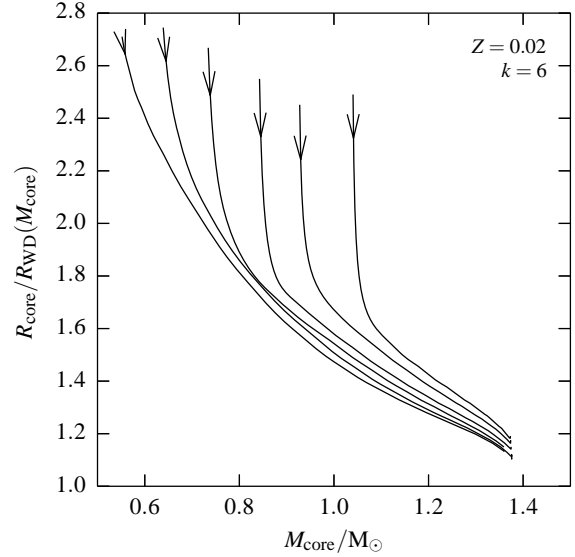}
  \caption{
The ratio of core radius to the radius of a cold white dwarf with mass equal to the core mass, for TPAGB stars with different zero-age masses of $Z=0.02$.
The masses are the same as in Fig.~\ref{fig:TPAGB-all}.
\textsc{bse} sets this ratio to be $5$.
}
  \label{fig:TPAGB-ratio}
\end{figure}

In \textsc{bse}, a TPAGB star has a core radius equal to five times the radius of a cold white dwarf with the same core mass.
Fig.~\ref{fig:TPAGB-all} shows the evolution in the core mass--core radius plane for $Z=0.02$ intermediate-mass stars in this phase and the core mass--core radius relation in \textsc{bse} TPAGB models.
For such stars, the degenerate core contracts as it increases in mass by hydrogen- and helium-shell burning.
The core radius is as well defined as for low-mass RGB stars.
The radii at $X=0.1$ and $0.001$ differ by no more than $7.5$ per cent.
The \textsc{bse} models have the right general behaviour -- decreasing core radius with increasing core mass -- but have core radii which are too large compared to the detailed models.

Fig.~\ref{fig:TPAGB-ratio} shows the ratio of core radius to white dwarf radius for a few TPAGB stars.
The \textsc{bse} formula overestimates the core radius by $185$ per cent at least.
Thus, we improve the fit in the same way as for low-mass HG and RGB stars.
For TPAGB stars, a $Z$-independent core mass--core radius relation is good enough to fit within $20$ per cent and improve on \textsc{bse}.
We choose a representative TPAGB model which covers the required range of core mass and radius and fit an improved polynomial fit to the ratio of core radius to white-dwarf radius at the same core mass.
The coefficients of the fit are given in \mbox{Table~\ref{tab:LM-TPAGB-ratio}} of Appendix A.
The new formula greatly improves on \textsc{bse}, giving the core radius at a given core mass, total mass and metallicity accurate to about $20$ per cent.
This is an important phase of evolution for the onset of CE evolution because the stars grow significantly.
It is therefore important to establish a better criterion for merging.

\subsection{Naked helium stars}
\begin{figure}
  \includegraphics[width=84mm]{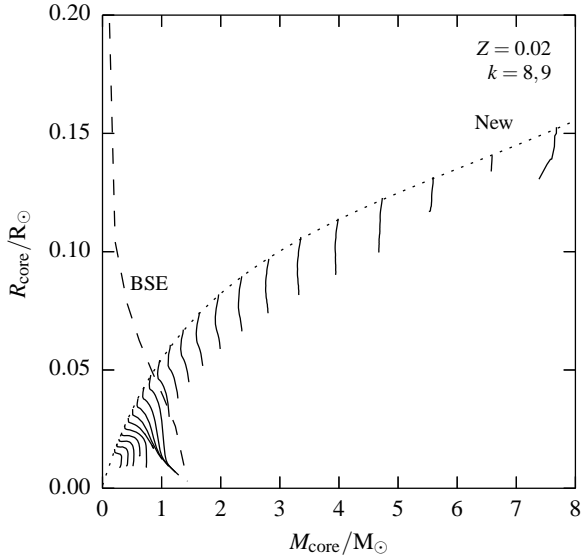}
  \caption{
Evolution of HeHG stars in the core mass--core radius plane.
The curves from left to right are for masses from $0.32$ to $10\,\rmn{M}_{\sun}$ in steps of $\Delta\log_{10}M = 0.0623$.
We compare the radii with the \textsc{bse} formulae for the core radius.
The \textsc{bse} algorithm gives a very poor estimate of the core radius, especially for helium stars which leave the MS with core masses greater than the Chandrasekhar mass.
We also show our new prescription for HeHG stars, the dot--dashed line.
}
  \label{fig:HeHG-all}
\end{figure}

In \textsc{bse}, the post-helium MS phase during which the helium-exhausted core of a naked helium star grows by shell burning is called the helium Hertzsprung gap (HeHG).
Naked helium stars with masses between about $0.9$ and $2.2\,\rmn{M}_{\sun}$ also undergo a red supergiant phase after the HeHG phase, referred to as the helium giant branch.
The \textsc{bse} prescription for the core radius is the same for both types of post-MS naked helium star: the core radius is five times the radius of a cold white dwarf with mass equal to the core mass.
The evolution ends either when the helium-rich envelope is nearly exhausted and the star becomes a CO white dwarf (COWD) without igniting carbon or the core mass grows to the Chandrasekhar mass or greater and a massless or core-collapse supernova occurs.

\subsubsection{Helium Hertzsprung gap}
\mbox{Fig.~\ref{fig:HeHG-all}} shows the evolution in the core mass--core radius plane for HeHG and giant branch stars.
For all masses, the core radius decreases as the core mass increases by helium-shell burning.
Also in the figure is the \textsc{bse} core mass--core radius relation which is immediately seen to be inadequate.
The relation is particularly poor for those stars with core masses exceeding the Chandrasekhar mass at the end of the helium MS.
According to \textsc{bse}, these stars have zero core radius.
We need to improve this prescription.
However, there is no single relation which provides a good fit in terms of core mass alone, a problem which is clearest at lower core masses where there are multiple core radii at the same core mass, and which is similar to that for the HG of hydrogen-rich stars (see Fig.~\ref{fig:RGB-intmass-all}).
The core mass and radius are not very well defined.
For example, in a $1.5\,\rmn{M}_{\sun}$, $Z=0.02$ helium star, the different definitions vary by about $0.1$--$0.5\,\rmn{M}_{\sun}$ in core mass at the same stage.
Similar variations are found for the other masses.

We improve the \textsc{bse} prescription for these stars by fitting to the core radius at the beginning of the HeHG as a function of core mass.
\textsc{bse} does not include a formula for the ZAMS radius for naked carbon stars so we cannot fit relative to this radius.
Instead, we fit a rational polynomial in $M_{\rmn{core}}$ to the core radius at the beginning of the HeHG.
We find a fit to better than $2$ per cent:
\begin{equation}
  \frac{R_{\rmn{core,HeHG}}}{\rmn{R}_{\sun}} = \frac{0.00123  + 0.0806 \mathcal{M}_{\rmn{core}} - 0.00331 \mathcal{M}_{\rmn{core}}^2}{1.00 + 0.467 \mathcal{M}_{\rmn{core}} - 0.0303 \mathcal{M}_{\rmn{core}}^2 },
\end{equation}
where $\mathcal{M}_{\rmn{core}} = M_{\rmn{core}} /\rmn{M}_{\sun}$.
This formula is shown in \mbox{Fig.~\ref{fig:HeHG-all}}.
It is more accurate than that of \textsc{bse} but has its own limitations.
If we were to set $R_{\rmn{core}} = R_{\rmn{core,HeHG}}(M_{\rmn{core}})$, then we would predict an increasing core radius with core mass at all masses.
The increase is particularly incorrect for the lower mass helium stars because then the core mass grows significantly during the HG phase.
To improve on this, we instead set $R_{\rmn{core}} = R_{\rmn{core,HeHG}}(M_{\rmn{core,BHG}})$ so that the core radius is constant after the helium MS and fixed by the core mass $M_{\rmn{core,BHG}}$ at the beginning of the HG phase.
This greatly improves on the \textsc{bse} prescription but gives core radii which are larger by up to about $100$ per cent compared to detailed models because the decrease in core radius is not modelled.
Further improvement is likely to require a fit including a dependence on total mass.

\subsubsection{Helium giant branch}
The formula given in the previous section also gives core radii in the helium giant-branch phase which are too large.
We improve on this by following the same procedure as for normal giant-branch stars with degenerate cores.
We find that a linear fit to the ratio of the core radius to the cold white dwarf radius works well,
\begin{equation}
  R_{\rmn{core,HeGB}} = \left(2.7 - 1.129 \mathcal{M}_{\rmn{core}} \right) R_{\rmn{WD}}(M_{\rmn{core}}),
\end{equation}
where $\mathcal{M}_{\rmn{core}} = M_{\rmn{core}} /\rmn{M}_{\sun}$.
This new formula gives core radii which are accurate to better than $20$ per cent in the later helium giant-branch phases.

\subsection{Summary}
We have compared the core radii of \textsc{bse} models with those in detailed models and found that \textsc{bse} requires some improvement in this respect.
Our new formulae (see Appendix A) greatly improve on \textsc{bse} for the RGB and AGB phases in the evolution of normal stars and the post-MS phases in the evolution of naked helium stars.
We have not modified the formulae for stars with non-degenerate helium cores, although we have discussed those phases in which the \textsc{bse} scheme is particularly inaccurate and made suggestions for how they might be improved in the future.
The \textsc{bse} formulae for these phases remain imperfect and should be used with caution: investigations which require the use of accurate core radii in these phases require detailed stellar models.

\section{Population synthesis}\label{sec:popsyn}
We compare properties of synthetic populations of binary star systems in which we include the formulae of the previous section and vary the critical radius used in the CE prescription.
Our aim here is to identify for which types of CE phase it is most important that the critical radius of the enveloper is correct.
To do this, we compute a simple model of the Galactic disc population of binary star systems.
Cataclysmic variables are used as an example of an important class of binary system formed through CE evolution.

\subsection{Model of the Galactic disc population}
Our basic model of the Galactic disc population is model A of \citet{Hur02:897}.
This model consists of a population of isolated binary systems each defined by its age $t$, its primary mass $M_1$, secondary mass $M_2$ and separation $a$ at zero age.
All the systems are assumed to begin their evolution with $Z=0.02$, circular orbits and with spins set according to the \citet{Hur02:897} prescription.

To compute statistical properties of the population we need the joint distribution of ages and zero-age parameters.
We write the fraction of systems with parameters in their respective infinitesimal ranges (e.g.,\ age in the range $t$ to $t+\rmn{d}t$) as $f_{\ln M_1, \ln M_2, \ln a, t}(M_1,  M_2, a, t)\,\rmn{d}\ln M_1\,\rmn{d}\ln M_2\,\rmn{d}\ln a\,\rmn{d}t$, where the probability density function (PDF)
\begin{equation}
  f_{\ln M_1, \ln M_2, \ln a, t}
=  f_{\ln M_1, \ln M_2}(M_1, M_2) \times  f_{\ln a}(a) \times f_t(t)
\end{equation}
and $f_x$ is the marginal PDF for the variable(s) $x$.
We thus make the usual assumption that the parameters are uncorrelated, except for the zero-age component masses.
The primary mass is distributed according to the \citet*{Kro93:545} initial mass function, $f_{M_1}\,\rmn{d}M_1 = \xi(M_1)\,\rmn{d}M_1$,
\begin{equation}
\xi \left( M \right) = \left\{ \begin{array} {l@{\quad}l} 
0 & \text{if } M/\rmn{M}_{\sun} \leq 0.1 \\ 
a_1 (M/\rmn{M}_{\sun})^{-1.3} & \text{if }0.1 < M/\rmn{M}_{\sun} \leq 0.5 \\ 
a_2 (M/\rmn{M}_{\sun})^{-2.2} & \text{if }0.5 < M/\rmn{M}_{\sun} \leq 1.0 \\ 
a_2 (M/\rmn{M}_{\sun})^{-2.7} & \text{if }1.0 < M/\rmn{M}_{\sun}
\end{array} \right. 
\end{equation}
where $a_1 = 0.290\,56$ and $a_2 = 0.155\,71$.
The zero-age mass ratio $q_2 = M_2/M_1$ obeys a flat distribution for $0<q_2<1$.
This implies that the joint PDF of logarithmic zero-age masses,
\begin{equation}
\begin{split}
  f_{\ln M_1, \ln M_2}(M_1, M_2) 
&= M_1 \xi(M_1) \times \left| \frac{\upartial (\ln M_1, q_2)}{\upartial (\ln M_1, \ln M_2)} \right|\\
&= \left\{ \begin{array}{l@{\quad}l} 
M_2 \xi(M_1) & \text{if } 0 < M_2 \leq M_1 \\ 
0 & \text{otherwise}, \\ 
\end{array} \right.
\end{split}
\end{equation}
where we have used the Jacobian determinant for the transformation from $(\ln M_1, q_2)$ to $(\ln M_1, \ln M_2)$.
The distribution of logarithmic zero-age separations is flat,
\begin{equation}
f_{\ln a}(a) = \left\{ \begin{array}{l@{\quad}l} 
0.123\,28 & \text{if } 3 < a/\rmn{R}_{\sun} \leq 10^4 \\ 
0 & \text{otherwise}. \\ 
\end{array} \right.
\end{equation}
The distribution of ages is the flat distribution appropriate to a constant star-formation history for a population with age $T$,
\begin{equation}
f_{t}(t) = \left\{ \begin{array}{l@{\quad}l} 
1/T & \text{if } 0 < t \leq T \\ 
0 & \text{otherwise}. \\ 
\end{array} \right.
\end{equation}

We classify CE phases according to the type of both stars at the onset and whether the phase ends by merging.
The expected rate of CE phases of such a given type $j$, when the age of the population is $T$, is
\begin{equation}
\begin{split}
  r_j
= & \frac{\rmn{d} N \langle \mathcal{N}_j\rangle}{\rmn{d} T} 
= \frac{\rmn{d}}{\rmn{d}T} \biggl[ N \int_D \mathcal{N}_j(M_{1}, M_{2}, a, t) \\ & \times f_{\ln M_1, \ln M_2, \ln a, t}\,\rmn{d}\ln M_{1}\,\rmn{d}\ln M_{2}\,\rmn{d}\ln a\,\rmn{d}t \biggr],
\end{split}
\end{equation}
where $\mathcal{N}_j(M_1, M_2, a, t)$ is the number of CE phases of type $j$ in the evolution of a system of age $t$ with zero-age parameters $M_1$, $M_2$ and $a$, the total number of systems is $N$ and the integral extends over all of parameter space $D$ described above.
For the given PDF, the expression simplifies to
\begin{equation}\label{eq:rate}
  r_j
= \frac{N}{T} \int_{D^\prime} \mathcal{N}_j(M_{1}, M_{2}, a, T) f_{\ln M_1, \ln M_2, \ln a}\,\rmn{d}\ln M_{1}\,\rmn{d}\ln M_{2}\,\rmn{d}\ln a.
\end{equation}
We set $T = 12\,\rmn{Gyr}$ and take the formation rate of binary systems $S = N/T = 7.6085\,\rmn{yr}^{-1}$.
We also compute the expected cumulative number of CE phases of type $j$:
\begin{equation}\label{eq:number}
  N \int_D \mathcal{N}_j(M_{1}, M_{2}, a, t) f_{\ln M_1, \ln M_2, \ln a, t}\,\rmn{d}\ln M_{1}\,\rmn{d}\ln M_{2}\,\rmn{d}\ln a\,\rmn{d}t.
\end{equation}

We compute the properties of three populations, binary systems evolved with the original \textsc{bse} prescription and those with our new prescription for the core radii for both choices of the critical radius.
We first compare the effect of using our new more accurate core radius formula with the pre-CE core radius set to be the CE critical radius, so that we can estimate some of the inaccuracies in studies that use the old formulae.
We then compare the effect of choosing the critical radius to be either the core radius at the onset of the CE phase or the radius of the post-CE remnant in \textsc{bse} with our new formulae.
For each population we compute the expected rate and cumulative number of CE phases classified by the pre-CE stellar types and whether the phase ends by merging.
We also compute the number of CVs which we define to be systems in which a white dwarf accretes matter from a Roche lobe filling companion.

\subsection{Numerical integration}
The integrals in equations~(\ref{eq:rate}) and (\ref{eq:number}) are computed with a generalization of the rectangle rule.
We divide the three-dimensional domain into a set of equal-sized cuboids with mid-points on a regular mesh between $M_i = 0.1$ and $80\,\rmn{M}_{\sun}$ for both masses, and between $a = 3$ and $10^4\,\rmn{R}_{\sun}$ in separation.
The mesh is defined by the combinations of $N_{\rmn{param}}$ evenly spaced points between the minimum and maximum of each logarithmic parameter.
For example, the meshpoint spacing for the component logarithmic masses is
\begin{equation}
  \Delta \ln M_{i} = (\ln M_{i,\rmn{max}} - \ln M_{i,\rmn{min}})/N_{\rmn{param}}
\end{equation}
where $M_{i,\rmn{min}}=0.1\,\rmn{M}_{\sun}$ and $M_{i,\rmn{max}}=80\,\rmn{M}_{\sun}$.
Thus, there are $N_{\rmn{param}}^2(N_{\rmn{param}}+1)/2$ cells, each with volume $V_{\rmn{cell}} = \Delta \ln M_{1} \times \Delta \ln M_2 \times \Delta \ln a$.
We use \textsc{bse} to quickly compute the evolution for each set of zero-age parameters and find $\mathcal{N}_j(M_1, M_2, a, T)$ for each type of CE phase.
Finally, the integral is approximated as a sum over the meshpoints where $\mathcal{N}_j$ is multiplied by the joint PDF at that meshpoint and the volume per cell.
For example,
\begin{equation}
  r_j = \sum_{a,b,c} \mathcal{N}_{j,a,b,c} \times f_{j,a,b,c} \times V_{\rmn{cell}},
\end{equation}
where the $(a,b,c)$ are meshpoint labels.
We use $N_{\rmn{param}} = 150$ for the populations discussed here and thus compute the evolution of about $1.7 \times 10^6$ binary systems for each population.
We compute $100$ populations for each choice of core radius formulae and CE critical radius.
We look for statistically significant differences in the mean rate of each type of CE phase by using a $t$-test, and taking $0.05$ as our critical $p$-value.
The integrals could also be computed with Monte Carlo methods but, unless the distribution of evolved stars is skewed towards interesting cases, many more systems must be evolved to achieve the same significance.

\section{Results and discussion}\label{sec:results}
We compare the populations computed with the new and old core radius formulae and then compare the populations in which the CE critical radius is chosen to be the pre-CE core radius or the stripped remnant radius.

\subsection{Core radius formulae}
\begin{table*}
\begin{minipage}{17.5cm}
\centering
\setlength{\tabcolsep}{2pt}
\caption{
Mean rates of CE merging in $\rmn{yr}^{-1}$ for populations computed with the \textsc{bse} core radius formulae.
The percentage changes in these rates when the new core radius formulae are used are given in parentheses: $(r_{\rmn{new}}/r_{\textsc{bse}} - 1)\times 100$.
The change is given only where there is a statistically significant difference in the mean rate.
The stellar types are defined in Table~\ref{tab:types}.
}
\label{tab:percentage-ircore}
\small
\begin{widetable}{17.5cm}{@{}llrlrlrlrlrlrlr@{}}
\toprule
  & \multicolumn{2}{c}{HG} &\multicolumn{2}{c}{RGB} & \multicolumn{2}{c}{CHeB} & \multicolumn{2}{c}{EAGB} & \multicolumn{2}{c}{TPAGB} & \multicolumn{2}{c}{HeHG} & \multicolumn{2}{c}{HeGB}  \\
\midrule
MSc & $1.1 \times 10^{-2}$ &  & $2.8 \times 10^{-2}$ &  & $3.4 \times 10^{-5}$ &  & $3.2 \times 10^{-4}$ & $(-0.14)$ & $1.3 \times 10^{-7}$ &  & $2.4 \times 10^{-3}$ &  & $0$ &   \\
MS & $4.8 \times 10^{-3}$ &  & $4.9 \times 10^{-2}$ &  & $7.6 \times 10^{-4}$ &  & $1.4 \times 10^{-3}$ &  & $2.0 \times 10^{-3}$ &  & $6.6 \times 10^{-4}$ &  & $4.1 \times 10^{-5}$ &   \\
HG & $8.2 \times 10^{-5}$ &  & $3.8 \times 10^{-3}$ & $(-33)$ & $1.4 \times 10^{-4}$ & $(-39)$ & $3.2 \times 10^{-4}$ &  & $4.8 \times 10^{-4}$ & $(-75)$ & $8.4 \times 10^{-4}$ & $(-45)$ & $3.3 \times 10^{-5}$ & $(-99)$  \\
RGB &    &  & $2.8 \times 10^{-3}$ & $(-95)$ & $4.6 \times 10^{-5}$ &  & $7.8 \times 10^{-5}$ &  & $6.7 \times 10^{-6}$ & $(-100)$ & $2.0 \times 10^{-4}$ & $(-1.9)$ & $0$ &   \\
CHeB &    &  &    &  & $3.8 \times 10^{-7}$ &  & $1.3 \times 10^{-3}$ &  & $0$ &  & $3.0 \times 10^{-5}$ &  & $2.2 \times 10^{-7}$ &   \\
EAGB &    &  &    &  &    &  & $1.7 \times 10^{-6}$ &  & $2.0 \times 10^{-7}$ &  & $6.8 \times 10^{-8}$ &  & $0$ &   \\
TPAGB &    &  &    &  &    &  &    &  & $0$ &  & $0$ &  & $0$ &   \\
HeMS & $2.2 \times 10^{-3}$ &  & $4.4 \times 10^{-7}$ & $(-100)$ & $2.6 \times 10^{-5}$ &  & $0$ &  & $0$ &  & $1.2 \times 10^{-3}$ & $(-0.57)$ & $0$ &   \\
HeHG &    &  &    &  &    &  &    &  &    &  & $3.0 \times 10^{-4}$ & $(-80)$ & $0$ &   \\
HeGB &    &  &    &  &    &  &    &  &    &  &    &  & $0$ &   \\
HeWD & $4.2 \times 10^{-3}$ & $(-41)$ & $4.3 \times 10^{-3}$ & $(-100)$ & $0$ &  & $0$ &  & $0$ &  & $1.6 \times 10^{-4}$ & $(-12)$ & $0$ &   \\
COWD & $4.3 \times 10^{-3}$ & $(-8.5)$ & $3.4 \times 10^{-4}$ & $(-100)$ & $5.1 \times 10^{-7}$ &  & $5.4 \times 10^{-5}$ & $(1.3)$ & $0$ &  & $3.7 \times 10^{-4}$ & $(-68)$ & $0$ &   \\
ONeWD & $9.5 \times 10^{-5}$ &  & $0$ &  & $4.1 \times 10^{-6}$ &  & $2.9 \times 10^{-5}$ &  & $0$ &  & $3.2 \times 10^{-5}$ & $(43)$ & $0$ &   \\
NS & $4.0 \times 10^{-4}$ &  & $3.8 \times 10^{-9}$ &  & $8.7 \times 10^{-7}$ &  & $1.0 \times 10^{-7}$ &  & $0$ &  & $1.9 \times 10^{-5}$ & $(66)$ & $0$ &   \\
BH & $6.7 \times 10^{-6}$ &  & $6.2 \times 10^{-10}$ &  & $0$ &  & $0$ &  & $0$ &  & $1.1 \times 10^{-5}$ &  & $0$ &   \\
\bottomrule
\end{widetable}
\end{minipage}
\end{table*}

There are important differences between the two populations when the new core radius formulae are used.
With the new formulae, the rate of merging in a CE is smaller by $10$ per cent and the total number of mergings in a CE is smaller by $7$ per cent.
Table~\ref{tab:percentage-ircore} gives the rates and percentage change in the rates as a function of the stellar types at the start of the CE phase.
The most significant differences are for CE phases of RGB stars with helium white dwarf (HeWD) or COWD companions, TPAGB+RGB CE phases and RGB+HeMS CE phases.
With the new formulae, no systems merge in CE phases of these types, whereas the corresponding systems do merge when we use the \textsc{bse} formulae.
The new formulae give smaller cores in general and thus CE phases initiated with the same pre-CE parameters can spiral to smaller separations, release more orbital energy and thus more easily eject the envelope and so avoid merging.
Consider the example of a system with zero-age parameters ($1.491\,\rmn{M}_{\sun}$, $0.534\,\rmn{M}_{\sun}$, $5.838\,\rmn{R}_{\sun}$), for which the evolution involves an RGB+HeWD CE phase.
The evolution with both formulae begins with the more massive component expanding off the MS and overfilling its Roche lobe as an MS star.
The system enters a phase of stable Roche lobe overflow which ends when the separation has increased to about $22\,\rmn{R}_{\sun}$ and the loser contracts to become a $0.190\,\rmn{M}_{\sun}$ helium-core white dwarf.
The initially less massive star now has a mass of $1.768\,\rmn{M}_{\sun}$ and expands to overfill its Roche lobe close to the base of the giant branch, when it is an RGB star with a core mass of $0.22\,\rmn{M}_{\sun}$.
A CE phase begins.
In the evolution with the \textsc{bse} formulae, the CE phase ends with the two components merging.
In the evolution with the new formulae, the envelope is successfully ejected to leave a binary system composed of two helium-core white dwarfs in a close orbit of $0.06\,\rmn{R}_{\sun}$.
Gravitational-wave radiation then tightens the orbit and the two components merge about $40\,\rmn{Myr}$ after the end of the CE phase.
The evolution with both formulae results in merging, but with the \textsc{bse} formulae this is in a CE and with the new formulae it is in a detached binary system after the end of the CE phase.
Thus, with the new formulae we change the results of studies which are concerned with the CE merging rates of RGB+HeWD systems or with the properties of the WD+WD population.
Specifically, the expected number of HeWD+HeWD systems is $1$ per cent larger with the new formulae.

This example demonstrates that, with the new formulae, the cases that just avoid merging usually go on to merge after the end of the CE phase because gravitational-wave radiation acts to shrink the orbit.
Aside from the fact that the improvements to the core radius formulae affect the populations of post-CE binary systems, it is still necessary to predict whether merging happens in a CE or soon after because some phenomena are thought to arise as a result of CE merging specifically.
For example, the carbon-rich K giants known as early-type R stars have been suggested to be the result of merging in RGB+HeWD CE phases \citep*{Izz07:661}.
We find no merging for CE phases of this type, and thus their conclusions should be revisited in light of this.

Table~\ref{tab:percentage-ircore} shows that there are differences for most other types of CE phase when the new formulae are used.
In most cases, there is a reduction in CE merging rates with the new formulae because the core radii had been mostly overestimated in \textsc{bse}.
This is not the case for HeHG stars, and we see from the table that there is an increase in the rate of merging for HeHG+NS and HeHG+ONeWD CE phases in particular.
This implies a decrease in the formation rate of NS/BH+NS systems.
The expected number of cataclysmic variables is not significantly affected by use of the new formulae.
With the new formulae the number of CVs is smaller by $0.19$ per cent.

\subsection{Critical radius}
\begin{table*}
\begin{minipage}{17.5cm}
\centering
\setlength{\tabcolsep}{2pt}
\caption{
Mean rates of CE merging in $\rmn{yr}^{-1}$ for populations computed with the new core radius formulae.
The percentage changes in these rates when the CE critical radius is changed are given in parentheses: $(r_{\rmn{stripped}}/r_{\rmn{core}} - 1)\times 100$.
The change is given only where there is a statistically significant difference in the mean rate.
The stellar types are defined in Table~\ref{tab:types}.
}
\label{tab:percentage-irce}
\small
\begin{widetable}{17.5cm}{@{}llrlrlrlrlrlrlr@{}}
\toprule
  & \multicolumn{2}{c}{HG} &\multicolumn{2}{c}{RGB} & \multicolumn{2}{c}{CHeB} & \multicolumn{2}{c}{EAGB} & \multicolumn{2}{c}{TPAGB} & \multicolumn{2}{c}{HeHG} & \multicolumn{2}{c}{HeGB}  \\
\midrule
MSc & $1.1 \times 10^{-2}$ &  & $2.8 \times 10^{-2}$ &  & $3.4 \times 10^{-5}$ &  & $3.2 \times 10^{-4}$ &  & $1.6 \times 10^{-7}$ &  & $2.4 \times 10^{-3}$ &  & $0$ &   \\
MS & $4.8 \times 10^{-3}$ &  & $4.9 \times 10^{-2}$ &  & $7.6 \times 10^{-4}$ &  & $1.4 \times 10^{-3}$ &  & $2.0 \times 10^{-3}$ &  & $6.6 \times 10^{-4}$ &  & $4.0 \times 10^{-5}$ &   \\
HG & $8.0 \times 10^{-5}$ &  & $2.5 \times 10^{-3}$ & $(-29)$ & $8.7 \times 10^{-5}$ &  & $3.1 \times 10^{-4}$ &  & $1.2 \times 10^{-4}$ & $(-8.9)$ & $4.7 \times 10^{-4}$ & $(-2)$ & $3.3 \times 10^{-7}$ &   \\
RGB &    &  & $1.5 \times 10^{-4}$ & $(-98)$ & $4.6 \times 10^{-5}$ &  & $7.7 \times 10^{-5}$ &  & $0$ &  & $2.0 \times 10^{-4}$ &  & $0$ &   \\
CHeB &    &  &    &  & $4.0 \times 10^{-7}$ &  & $1.3 \times 10^{-3}$ &  & $0$ &  & $2.9 \times 10^{-5}$ &  & $2.3 \times 10^{-7}$ &   \\
EAGB &    &  &    &  &    &  & $1.9 \times 10^{-6}$ &  & $1.7 \times 10^{-7}$ &  & $0$ &  & $0$ &   \\
TPAGB &    &  &    &  &    &  &    &  & $0$ &  & $0$ &  & $0$ &   \\
HeMS & $2.2 \times 10^{-3}$ &  & $0$ &  & $2.6 \times 10^{-5}$ &  & $0$ &  & $0$ &  & $1.2 \times 10^{-3}$ &  & $0$ &   \\
HeHG &    &  &    &  &    &  &    &  &    &  & $6.0 \times 10^{-5}$ &  & $0$ &   \\
HeGB &    &  &    &  &    &  &    &  &    &  &    &  & $0$ &   \\
HeWD & $2.5 \times 10^{-3}$ & $(-60)$ & $1.7 \times 10^{-7}$ & $(-100)$ & $0$ &  & $0$ &  & $0$ &  & $1.4 \times 10^{-4}$ &  & $0$ &   \\
COWD & $4.0 \times 10^{-3}$ & $(-10)$ & $0$ &  & $5.1 \times 10^{-7}$ &  & $5.5 \times 10^{-5}$ &  & $0$ &  & $1.2 \times 10^{-4}$ &  & $0$ &   \\
ONeWD & $9.5 \times 10^{-5}$ &  & $0$ &  & $4.1 \times 10^{-6}$ &  & $2.9 \times 10^{-5}$ &  & $0$ &  & $4.5 \times 10^{-5}$ & $(1.2)$ & $0$ &   \\
NS & $4.0 \times 10^{-4}$ &  & $8.1 \times 10^{-10}$ &  & $8.1 \times 10^{-7}$ & $(11)$ & $8.5 \times 10^{-8}$ &  & $0$ &  & $3.1 \times 10^{-5}$ &  & $0$ &   \\
BH & $6.8 \times 10^{-6}$ &  & $6.5 \times 10^{-10}$ &  & $0$ &  & $0$ &  & $0$ &  & $1.1 \times 10^{-5}$ &  & $0$ &   \\
\bottomrule
\end{widetable}
\end{minipage}
\end{table*}

We now compare populations in which the critical radius in CE evolution is taken to be the radius of the completely stripped remnant at the end of the CE phase or the pre-CE core radius.
We use our new formulae for core radii in both populations.

The current version of \textsc{bse} requires the critical radius to be the pre-CE core radius for naked helium stars because it does not compute the evolution of (nearly) naked carbon stars.
In reality we expect a helium CE phase to produce a nearly naked carbon star which burns carbon before core collapse.
Instead, \textsc{bse} sets the remnant of a helium CE phase to be the eventual core-collapse remnant, a neutron star or black hole.
For this reason, our improvements to the core radius formulae for naked helium stars are important because this is the only critical radius available for CE phases involving these objects.
Also, as discussed in Section~\ref{sec:intmass}, the stripped remnant radius and the pre-CE core radius in \textsc{bse} are equal for stars with non-degenerate helium cores.
Therefore we only expect variations of the critical radius to affect CE phases involving low-mass HG, RGB and TPAGB stars.

Table~\ref{tab:percentage-irce} compares the rates in the two populations and shows that there are differences for these cases.
The largest percentage differences are for the rates of RGB+RGB and RGB+HeWD CE merging which is $98$ per cent smaller when the critical radius is chosen to be the stripped remnant radius.
This implies that the formation rate of HeWD+HeWD systems is larger when the critical radius is changed to the stripped remnant radius.
The change in the critical radius makes the total rate of merging in a CE smaller by $2$~per~cent and the total number of mergings in a CE smaller by $2$~per~cent.
The expected number of CVs is not significantly affected by changing the CE critical radius.
When the critical radius is changed to the stripped remnant radius, the number of CVs is larger by $0.06$ per cent.
Many of the most interesting post-CE systems are fortunately unaffected by both our improvements to the core radius formulae and the additional theoretical complication of the uncertain critical radius.
These include the RGB or AGB+MS CE phases which have been the focus of previous studies of post-CE systems \citep*{Dav10:179} and merging \citep{Pol10:1752}.
However, we have shown that the uncertainty affects predictions relating to important systems, such as double white dwarfs, formed after a CE is successfully ejected, as well as objects produced by merging in a CE such as the early R-type stars.
If studies of CE merging such as that by \citet{Pol10:1752} are extended to CE phases in more evolved systems, then this additional uncertainty should be considered.

Ultimately, it is difficult to constrain what the critical radius should be in the cases that matter, because $\alpha$ can be varied, independently of how well known the core radii are.
We have presented results for $\alpha=3$ here, but our new formulae improve the accuracy of populations whatever $\alpha$ is used.
Because space densities of particular classes of systems are very poorly known, it is hard to make significant comparisons with observations.

\section{Conclusion}\label{sec:conclusion}
The \textsc{bse} algorithms \citep{Hur02:897} and various formulae from them are widely used for synthesis of populations of binary star systems.
Many approximations and educated guesses were made in their contribution.
Here, we look more closely at two related weak points -- the critical radius chosen for predicting the outcome of CE evolution and the formulae for the core radii of stars.
We have derived greatly improved formulae for the core radii and find that this significantly changes the rates of merging in CE phases involving RGB stars, when the critical radius is the pre-CE core radius.
The types of CE merging which are significantly affected by our improvements to the core radius formulae are those with large absolute percentages in Table~\ref{tab:percentage-ircore}.
Using the original \textsc{bse} formulae can lead to considerable errors in the modelled rates for merging in CE phases and for important post-CE systems such as double white dwarfs.
We have compared the choice of either the radius of the naked remnant (e.g.,\ cold white dwarf radius) or the pre-CE core radius with the end-of-CE Roche lobe.
Our improvements to the core radii have a much larger effect on the rates of CE phases of different types than the effect of changing the critical radius.
The types of CE merging which are significantly affected by variations in the CE critical radius are those with large absolute numbers in Table~\ref{tab:percentage-irce}.
This additional uncertainty in the treatment of CE evolution is also relevant to users of other rapid evolution and population synthesis codes which are based on the single-star formulae of \citet{Hur00:543}.
While the $\alpha$ prescription continues to be relied upon, future investigations should be aware of the uncertainty we have noted here and how it relates to a detailed model of mass transfer.
Those using the \textsc{bse} formulae would be better served by our new formulae for the core radius.

\section*{Acknowledgements}
PDH thanks the Science and Technology Facilities Council, STFC, for his studentship. 
CAT thanks Churchill College for his fellowship and all it entails.

\balance
\bibliographystyle{mn2e}
\bibliography{phd}

\onecolumn
\appendix
\section{Coefficients for fits}
The core radius of a low-mass RGB or TPAGB star with metallicity $Z$ and core mass $M_{\rmn{core}} = \mathcal{M}\,\rmn{M}_{\sun}$ is
\begin{equation}\label{eq:coeff-1}
  \frac{R_{\rmn{core}}}{\rmn{R}_{\sun}} = 
\left(c_0 + 
c_1 \mathcal{M} + 
c_2 \mathcal{M}^2 + 
c_3 \mathcal{M}^3\right) R_{\rmn{WD}}(M_{\rmn{core}})
\end{equation}
where 
\begin{equation}\label{eq:coeff-2}
c_n = 
\alpha + 
\beta \zeta + 
\gamma {\zeta}^2 + 
\eta {\zeta}^3 +
\mu {\zeta}^4 +
\theta {\zeta}^5,
\end{equation}
$\zeta = \log_{10}(Z/0.02)$, $R_{\rmn{WD}}$ is given by equation~(\ref{eq:RWD}) and the coefficients are given in Tables \ref{tab:LM-HG-RGB-ratio} and \ref{tab:LM-TPAGB-ratio}.
Note that, in the tables, $x(n)$ represents $x \times 10^n$.
The coefficients for TPAGB stars do not depend on $Z$.

\centering
\captionof{table}{
Coefficients in equations (\ref{eq:coeff-1}) and (\ref{eq:coeff-2}) for low-mass HG and RGB stars.
}
\label{tab:LM-HG-RGB-ratio}
\small
\begin{tabular}{lllllll}
\toprule
  & \multicolumn{1}{c}{$\alpha$} & \multicolumn{1}{c}{$\beta$} & \multicolumn{1}{c}{$\gamma$} & \multicolumn{1}{c}{$\eta$} & \multicolumn{1}{c}{$\mu$} & \multicolumn{1}{c}{$\theta$} \\
\midrule
 $c_0$  & $\phantom{-}2.817859(+0)$ & $\phantom{-}4.331671(-1)$ & $-8.152041(-1)$ & $-1.329429(+0)$ & $-3.502317(-1)$ & $\phantom{-}0$ \\
 $c_1$  & $-1.454750(+1)$ & $-3.844703(+0)$ & $\phantom{-}8.279414(+0)$ & $\phantom{-}1.252224(+1)$ & $\phantom{-}3.256815(+0)$ & $\phantom{-}0$ \\
 $c_2$  & $\phantom{-}4.947425(+1)$ & $\phantom{-}1.369688(+1)$ & $-2.513380(+1)$ & $-3.773333(+1)$ & $-9.791961(+0)$ & $\phantom{-}0$ \\
 $c_3$  & $-4.914713(+1)$ & $-1.326969(+1)$ & $\phantom{-}4.137217(+1)$ & $\phantom{-}6.525871(+1)$ & $\phantom{-}2.617208(+1)$ & $\phantom{-}3.184125(+0)$ \\
\bottomrule
\end{tabular}

\captionof{table}{
Coefficients in equation (\ref{eq:coeff-1}) for TPAGB stars.
}
\label{tab:LM-TPAGB-ratio}
\small
\begin{tabular}{ll}
\toprule
        & \multicolumn{1}{c}{$\alpha$} \\
\midrule
 $c_0$  & $\phantom{-}7.342054(+0)$ \\
 $c_1$  & $-1.328317(+1)$ \\
 $c_2$  & $\phantom{-}1.020264(+1)$ \\
 $c_3$  & $-2.786524(+0)$ \\
\bottomrule
\end{tabular}

\label{lastpage}

\end{document}